\documentclass[%
 reprint,
 amsmath,amssymb,
 aps,
10pt]{revtex4-1}
\usepackage{xcolor}
\usepackage{graphicx}
\usepackage{dcolumn}
\usepackage{bm}

\begin{document}


\title{Spectral Walls in Soliton Collisions}

\author{C. Adam}
 \affiliation{Departamento de F\'isica de Part\'iculas, Universidad de Santiago de Compostela and Instituto Galego de F\'isica de Altas Enerxias (IGFAE) E-15782 Santiago de Compostela, Spain.}
\author{K. Oles}%

\author{T. Romanczukiewicz}

\author{A. Wereszczynski}

\affiliation{
 Institute of Physics,  Jagiellonian University, Lojasiewicza 11, Krak\'{o}w, Poland
}

\date{\today}

\begin{abstract}
During defect-antidefect scattering, bound modes frequently disappear into the continuous spectrum before the defects themselves collide. This leads to a structural, nonperturbative change in the spectrum of small excitations. Sometimes the effect can be seen as a hard wall from which the defect can bounce off. We show the existence of these \textit{spectral walls} and study their properties in the $\phi^4$ model with BPS preserving impurity, where the spectral wall phenomenon can be isolated because the static force between the antikink and the impurity vanishes. We conclude that such spectral walls should surround all solitons possessing internal modes. 
\end{abstract}

\pacs{Valid PACS appear here}
\maketitle


{\bf Motivation: \;} 
The detailed understanding of soliton interactions in non-integrable models is a difficult and only partially resolved problem. 
The prototypical $\phi^4$ model in (1+1) dimensions, for example, reveals kink-antikink collisions with 
a chaotic structure, typically associated with the existence of one or several internal modes which may be excited during the scattering process
 \cite{Sugiyama:1979mi, Weigel:2018xng, Campbell:1983xu, Anninos:1991un, Goodman2005hbr}.
These modes can store energy, binding the solitons for a while, and may eventually transfer their energy back to the translational degrees of freedom. 
This energy exchange has many properties of a resonant phenomenon, leading to an intriguing fractal-like pattern of multiple
 bounce windows as a function of the initial velocity. A similar mechanism was observed in other solitonic models, including 
 soliton-impurity collisions \cite{Zhang1992, Fei:1992dk}, multi-component fields \cite{Halavanau:2012dv, Alonso-Izquierdo:2017zta} and quasinormal modes \cite{Dorey:2017dsn}. 
 A bound oscillational mode can even be excited when there are no such modes on the original solitons (asymptotic states). This may happen, for example, in a model with two vacua with two different mass parameters \cite{Dorey:2011yw}.

An effective model was proposed in \cite{Sugiyama:1979mi} and later studied and modified by many others, where the relative position, $a$, of the colliding defects and the amplitude of the bound (vibrational) mode, $A$, were introduced as collective coordinates.  Initially, this model seemed to confirm the above results 
quite well.  Recently, however, a typographical error in \cite{Sugiyama:1979mi} was corrected \cite{Weigel:2013kwa, Takyi:2016tnc}, and the resulting picture did not correctly describe the multi-bounce windows characteristic for soliton 
collisions.
As pointed out in \cite{Weigel:2018xng}, one reason is that all analytical attempts assumed that the field can be written as a 
simple superposition of the kink, antikink and mode profiles. 

However, this picture is not correct, in general, because of several related problems: 
\\
$\bullet$ The separated kink-antikink pair is not a static solution, therefore the solitons are deformed by mutual interactions (static forces) before they collide,
\\
$\bullet$ as none of the intermediate states is static, the eigen-problem for the bound modes is not well-defined,
\\
$\bullet$ at the moment of the collision $a=0$ the solitons vanish, which is identified as a zero vector problem,  
\\
$\bullet$ when solitons momentarily vanish, there can be no bound states. 

\begin{figure}
\includegraphics[width=1.00\columnwidth]{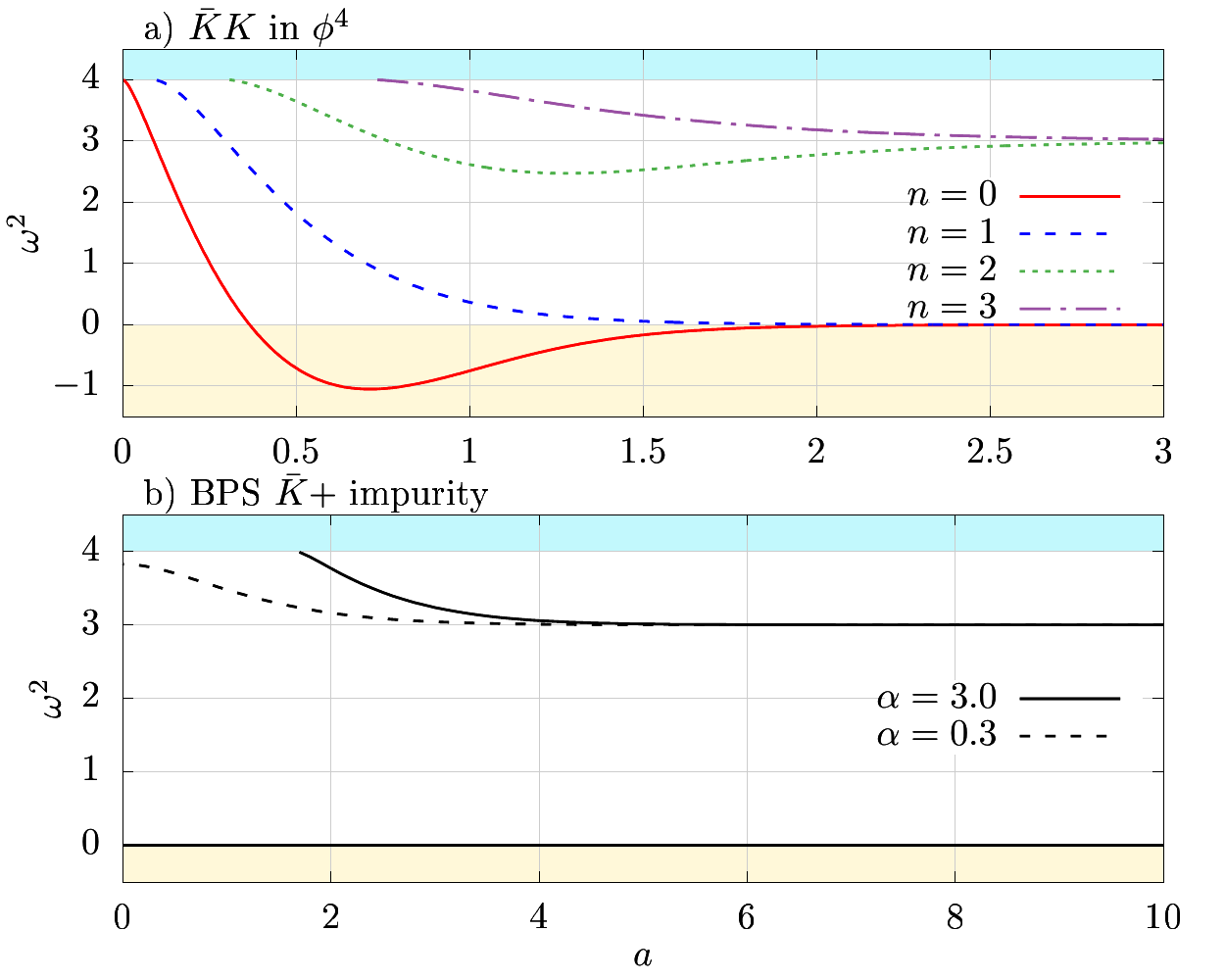}
 \caption{(a) Spectral structure of the $K\bar{K}$ superposition in $\phi^4$. (b)  Spectral structure of the static BPS $\bar K$ solution in the BPS-impurity model  for two values of $\alpha$. }\label{fig:spectralphi4}
\end{figure}

Actually, the bound modes vanish even earlier, see, e.g.,  Fig. \ref{fig:spectralphi4}, upper panel, 
where we plot the spectral structure of the kink-antikink configuration \mbox{$\phi^0 =\tanh(x-a)-\tanh(x+a)+1$} in the $\phi^4$ model. This configuration is not a solution, in general, but it {\em is} a solution in the limits $a\to \pm \infty$ and $a\to 0$. Fig. \ref{fig:spectralphi4}a), therefore, demonstrates that the modes {\em must} change and, in particular, disappear into the continuum as the solitons approach each other.

Thus, the collective coordinate dynamics does not correspond to the real dynamics of the process. 
Even beyond the effective model,  the mixing of the the kink-mode interaction with the (static) forces between solitons, which change the soliton profiles and their spectral properties,  renders any analytical treatment very difficult.
 
This mixing problem could be avoided for a theory with static multi-soliton solutions of the Bogomolny-Prasad-Sommerfield (BPS) type. Then, individual solitons of a static multi-soliton configuration do not interact, like, e.g., the vortices in the Abelian Higgs model at critical coupling. These vortices can be placed at arbitrary positions, leading to a finite-dimensional moduli space. 
Different multi-vortex configurations on this moduli space, however, have in general different overall profiles (shapes) and, therefore, they vibrate differently, i.e., their spectral structures differ. 
The low energy scattering of BPS solitons can be described as a geodesic motion on moduli space. In a next step, a mode on a scattered soliton can be excited. In this way, one could disentangle the soliton-mode interaction from the inter-soliton forces. 

In soliton models in (1+1) dimensions, however, only one-soliton solutions belong to the BPS sector. The corresponding moduli space is trivial and given by translations of the kink. In particular, the spectral structure remains unchanged along this very simple moduli space. 

Very recently it has been observed that this situation may change when an {\em impurity} is added. In particular, there exist BPS-impurity models in (1+1) dimensions \cite{Adam:2018pvd, Adam:2018tnv} whose moduli spaces resemble the higher-dimensional cases in that {\it the spectral structure of the soliton-impurity solution depends on its position on moduli space}. This gives us the unique opportunity to disentangle the above-mentioned mixing between the kink-mode interaction and the inter-kink force. 

The aim of this letter is to analyze the interaction of the excited mode with the BPS soliton in the BPS-impurity $\phi^4$ theory. We discover a universal phenomenon, {\it a spectral wall}, which denotes a {\it spatially localised region}, defined by the point where an oscillation mode enters the continuous spectrum. At this point, a
 nontrivial modification of the soliton interaction occurs. In the simplest
case, it is just a hard-wall reflection, but other, more complicated
 patterns are possible, depending on the modes carried by the incoming
 soliton.


{\bf BPS-impurity model: \;}
The BPS-impurity model is defined by the following Lagrangian 
\begin{equation}\label{eq:lag1}
 L=\int dx \left[ \frac{1}{2} \phi_t^2 - \frac{1}{2} \left( \phi_x +\sqrt{2} W + \sqrt{2} \sigma \right)^2 \right]
 \end{equation}
 where $\phi (t,x)$ is a scalar field and $W=W(\phi)$ a prepotential such that the potential of the original model without impurity is $U(\phi)=W^2$. Finally, $\sigma=\sigma(x)$ is a spatially localized impurity. The model is a half-BPS theory in the sense that half of the solitons (here the antikinks) of the no-impurity field theory remain BPS solitons, that is, saturate a pertinent topological bound and obey the corresponding Bogomolny (first order static differential) equation. 
Furthermore, the BPS sector also contains  topologically trivial solutions (lumps) which are the counterparts of the vacuum solutions $\phi = \phi_v=$ const. (minima of $U$) of the model without impurity. 
The trivial moduli space (spatial translations) of the model without impurity now transforms into a nontrivial  one-dimensional
 moduli space $\mathcal{M}$ of generalized translations in the BPS sector \cite{Adam:2018tnv}, \cite{Adam:2019yst}. It can be parametrized, e.g.,  by a point $a \in \mathbb{R}$ measuring the distance between the static BPS soliton and the impurity (no static force between them). Concretely, we locate the impurity at $x=0$ and choose for $a$ the position $x=a$ where the field of the antikink vanishes.
  
In the present work, we analyze the BPS-impurity version of the $\phi^4$ model. Hence, we assume \mbox{$W=(1-\phi^2)/\sqrt{2}$}. Further, we choose the impurity \mbox{$\sigma =\alpha/\cosh^2 x$} \cite{Adam:2018tnv} ($\alpha$ is a real parameter which measures the strength of the impurity). 
Owing to the generalized translational symmetry, the Bogomolny equation
\begin{equation}
\frac{1}{\sqrt{2}}\phi_x+W(\phi)+\sigma (x)=0
\end{equation}
is solved by a one-parameter family of BPS antikink solutions $\phi^0(x;a)$.
Therefore, in the BPS-impurity model the spectral structure is well-defined for any $a$. 
The modes can be found by a small perturbation around the static BPS solution $\phi = \phi^0(x;a) + A \eta (x,t; a) e^{i\omega t}$
\begin{equation}
- \eta_{xx} + 2\left[ W_\phi^2 + W_{\phi \phi}\left( W_\phi +\sigma \right) \right] \eta= \omega^2 \eta
\end{equation}
where the mode amplitude $A$ is assumed to be small. Further, $W$ and its derivatives are calculated for $\phi=\phi^0$. 
Obviously, the spectral structure depends on $a$, i.e., on the position on moduli space.

In this letter, we consider two examples of the simplest case with one discrete mode (plus one zero mode for the generalized translational symmetry). In the first example $\alpha = 0.3$, the mode exists for all points on the moduli space and its frequency grows while we approach the impurity, see Fig. \ref{fig:spectralphi4} (lower panel, dotted curve). In the second example $\alpha = 3$, we choose the impurity such that, at a certain point on the moduli space (a certain critical distance between the BPS antikink and the impurity)  the mode enters the continuous spectrum, becoming a quasinormal mode, see Fig. \ref{fig:spectralphi4} (lower panel, solid curve). Here it happens for $a=a_{\rm cr}=\pm 1.68$. Obviously, there is a fundamental qualitative difference between the two cases. While for $\alpha=0.3$ we expect a smooth evolution (captured by an effective model), for  $\alpha=3$ we expect some novel effects.

Indeed, as we will see below, this drastic change in the spectrum of discrete modes leads to the appearance of a \textit{spectral wall}, i.e., a spatially well-localized region (barrier) at which the BPS antikink with the pertinent mode excited may bounce back or be trapped, even though the unexcited BPS antikink goes through this point without any interference (no energy loss due to the geodesic motion on moduli space).

Of course, the spectral structure can reveal even more complicated patterns with a bigger number of modes entering the continuous spectrum at different points. This may lead to more involved structures and new effects which, however, we leave for future investigations.

\begin{figure}
\centering
 \includegraphics[width=1\columnwidth]{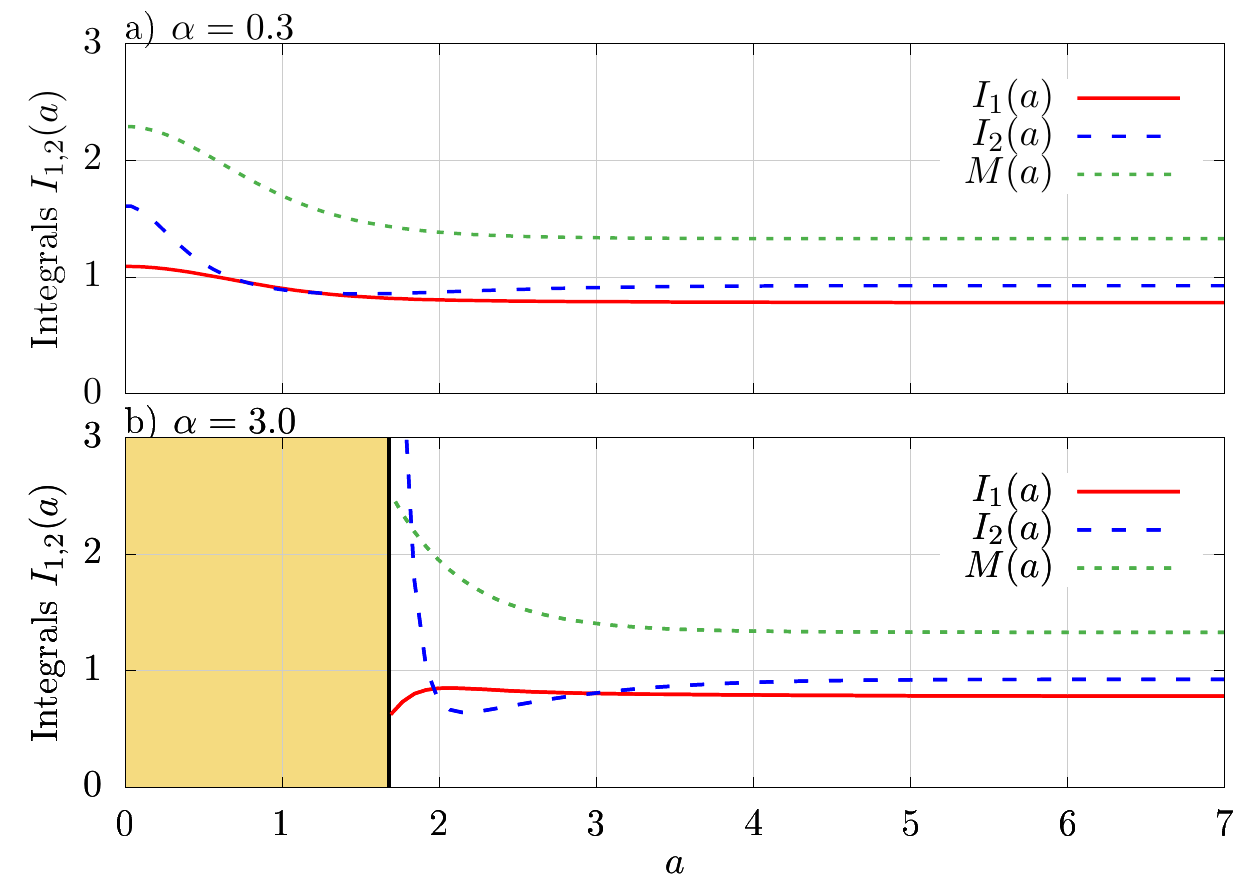}
 \caption{Regular (a) and singular (b) terms in the lagrangian of the effective model. The singularities in (b) occur at the spectral wall $a_{\rm cr} = \pm 1.68$.}\label{fig:Integrals}
\end{figure}

{\bf Effective model: \; }
 The standard collective coordinate method (CCM) consists in expressing the field as a superposition of known profiles such as kinks and their bound modes \cite{Sugiyama:1979mi}, leaving the positions and mode excitations as the only dynamical variables (collective coordinates). As explained in the introduction, 
this approach has important problems, in general. Let us now consider the CCM for the BPS model. It corresponds to taking a field of the form 
\begin{equation}\label{eq:BPSform}
 \phi(x,t)=\phi^0(x;a(t))+A(t)\eta(x;a(t)).
\end{equation} 
Some problems of the CCM are avoided in the BPS model, 
because both the static solution and the spectral structure are well-defined for all $a$. 
Inserting the above expression into the lagrangian, we get at quadratic order in $a$ and $A$
\begin{equation} \label{eff-mod}
L=\frac{1}{2} \dot{A}^2 + \frac{1}{2} I_1A^2 \dot{a}^2  +\frac{1}{2}M\dot{a}^2 + A \dot{a}^2 I_2-\frac{1}{2} \omega^2 A^2
\end{equation}
where 
\begin{equation}
 M = \int (\phi_a^0)^2\,,\quad I_1=\int \eta_a^2\,,\quad I_2=\int \eta_a \phi_a^0. 
\end{equation} 
$M(a)$ is the effective mass of the BPS soliton or a metric on moduli space and, therefore, is well-defined for any position on $\mathcal{M}$. However, the two other integrals can be divergent as $a$ approaches a critical separation $a_\text{cr}$, simply because the mode becomes non-normalizable at $\omega=2$, when it enters the continuum (Fig. \ref{fig:Integrals}). Obviously, at this point the effective model (\ref{eff-mod}) must break down and some new effects are expected. 
So, while the CCM breaks down also in the BPS case for some parameter values, it contains all the relevant information about its possible range of validity and the points in parameter space where its break-down occurs. Even within its range of validity, the CCM will provide quantitatively reliable results only for a sufficiently slow (adiabatic) evolution, such that additional degrees of freedom (not included in the CCM) are not excited too much in the full time evolution. We shall see below to which degree this adiabatic condition can be met in the BPS model. 

One way to understand the problem of the mode entering the continuum is that two different sets of independent variables are required on both sides of $a_\text{cr}$, and there is no obvious way to match effective models in both regions. Moreover, the bound mode entering the continuous spectrum becomes a quasi-normal mode. In \cite{Dorey:2017dsn} it was shown that such modes can also be responsible for creating a resonance structure. But constructing an effective model with a quasinormal mode is not as straightforward as in the case of bound modes.

%

\begin{figure}
 \includegraphics[width=1.0\columnwidth]{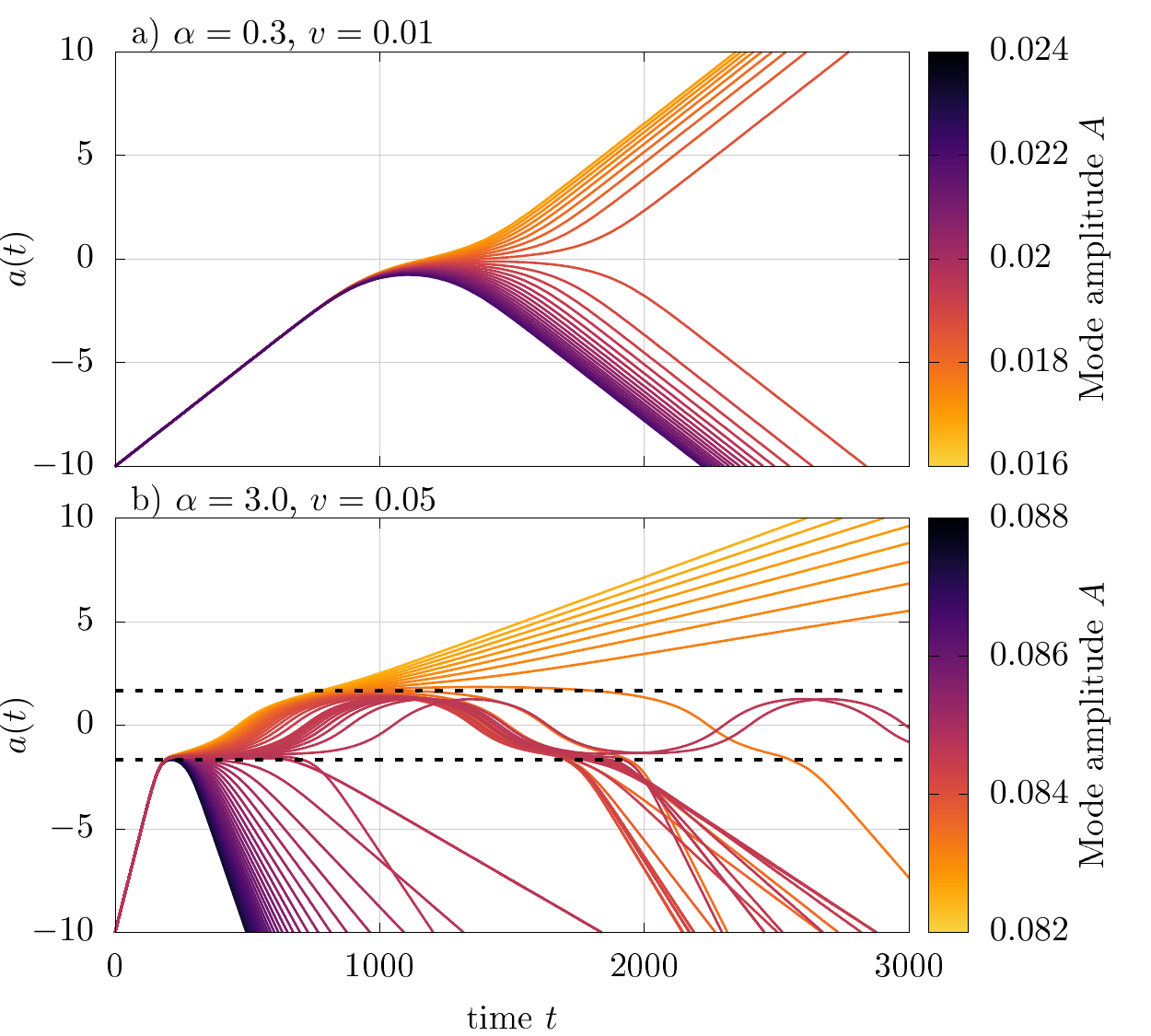} 
 \caption{Comparison between the smooth evolution for $\alpha=0.3$ and $v=0.01$ (a) and a meandering kink trapped between spectral walls for $\alpha=3$ and $v=0.05$ (b)  for different excitations of the mode. Dashed lines correspond to the positions of the spectral walls.}\label{fig:numericalPassages}
\end{figure}

{\bf The spectral wall: \; }
We have collided the BPS antikink initially separated by $a(0)=-10$ with the impurity for $\alpha=0.3$ and $\alpha=3.0$ (Fig. \ref{fig:numericalPassages}; in fact, we have considered many more values of $\alpha$, all leading to the same results. For simplicity, we only present two values, covering the two generic cases of qualitatively different behavior). In both cases, the unexcited solitons smoothly travel through the impurity, as expected from the geodesic flow on moduli space. 
However, when the internal mode of the antikink is excited, we have found that the soliton can bounce back for small velocities $v=\dot{a}(0)$. For $\alpha = 0.3$, the position of the turning point is changing smoothly both with the amplitude of the excitation and with the velocity reaching the origin.

The situation in the $\alpha=3$ case is different. Increasing the amplitude of the excitation for fixed velocity, the soliton is slowing down at the position of the spectral wall $a_\textrm{cr}=-1.68$. If the excitation is large enough, the antikink bounces from the wall. Decreasing the amplitude slightly we have found another intriguing effect. The soliton can go through the wall but is reflected from the second (symmetric) wall  behind the impurity. Sometimes even a few internal reflections can be observed. When the perturbation of the antikink is radiated out, it can finally pass one of the walls. Effectively there is a small window in the amplitude range, inside which the soliton can bounce back or go through the impurity after a series of internal reflections. This suggests that, even after a long time, the energy stored in the mode is still attached to the kink in some way, and it takes time to radiate it out. Indeed, after entering the continuous spectrum, the normal mode attached to the antikink turns into a quasinormal mode whose frequency and width increase and admit the highest value $\omega=3.72+0.11i$ for $a=0$ (determined using Prone's method). The existence of such a mode can prevent the immediate emission of energy. 
We also have found that the condition for the bounce, for small velocity, obeys a linear scaling law $A\approx 1.70 v$.

We have compared the results of the numerical simulation of the full PDE problem with predictions from the effective model (\ref{eff-mod}). 
In the $\alpha=0.3$ case, when the effective model is applicable for all separations, we have found a very good agreement for small initial velocities, implying that the evolution is adiabatic, i.e., the velocity and acceleration are sufficiently small during the whole time evolution.  For example, for an initial velocity $v=0.01$, the critical excitation $A(0)$ separating bounce from passage agrees with an accuracy of about 1\% with the full numerics ($A_\text{full}=0.0186$ vs. $A_\text{eff}=0.0188$). Even for $\alpha=3$, for low velocities the effective model works well until the the antikink gets close to the wall. For $v=0.05$, e.g., we found a critical excitation $A_\text{full}=0.0847$ and  $A_\text{eff}=0.0878$, which is slightly less accurate, but the integrals calculated near the wall have larger numerical errors.

Higher, relativistic velocities require a higher mode excitation. However, in nonlinear models the frequency of the highly excited mode depends on the amplitude, and usually is lower than the eigenfrequency found from the linearization. The frequency shift means that the highly excited mode  enters the continuous spectrum for smaller values of $a$. This effect was also observed numerically, when $A_{cr}>0.2$ the wall shifts visibly.
This may be important in the more general, non-BPS case, because the colliding objects can attract each other, therefore there should exist some minimal $A_{cr}$ below which the full collision would take place even with zero initial velocity. Note that in the pure $\phi^4$ model the resonant structure can be observed for relativistic velocities $0.18<v<0.26$ measured in the center of mass. Such high velocity collisions mean that both the inter-soliton interaction and the mode excitation are very large.

A similar wall was also found when the system undergoes a transition from three to two oscillating bound modes ($\alpha<-0.35$).

{\bf Summary: \; }
In this letter, we investigated in detail how soliton scattering is affected by the interaction of the colliding solitons with an internal, vibrational mode, in particular, when this mode disappears into the continuum.
For this purpose, we chose the simplest possible setting, the BPS-impurity model, which allows to {\em isolate} the soliton-mode interaction, because, owing to the BPS property, static inter-soliton forces are absent. We found that, at the point where the mode crosses into the continuum, the dynamics of the scattering experiences a drastic modification, in the simplest case a hard-wall reflection ("spectral wall").
We expect the spectral wall to be a generic phenomenon for the interaction of solitons in non-integrable theories, if a discrete mode undergoes a transition to the continuous spectrum. Of course, its effects might be less visible than in the BPS-impurity model, because other interactions may interfere. 
In the case of solitons with long-range tails, e.g., it is practically impossible to find an unperturbed initial state, therefore all collisions in such systems are collisions of excited states \cite{Christov:2018wsa}.  Furthermore, in more realistic physical systems solitons are always excited due to quantum or thermal fluctuations. 

The spectral wall should be especially easy to find in BPS theories (if a mode transition occurs) like the Abelian Higgs model in (2+1) dimensions at critical coupling. Our results, therefore, provide new insights into the dynamics of BPS solitons beyond the geodesic approximation, where the spectral wall effect may play a significant role.  

Moreover, we show that the disappearance of the bound modes may be responsible for the failure to construct a reliable collective coordinate model for kink-antikink collision processes in theories like the $\phi^4$ model. 

Finally, we remark that kink-impurity scattering in a $\phi^4$ model coupled to a $\delta$ function impurity (in a non-BPS preserving way) was studied in \cite{Zhang1992}, both numerically and within the CCM. In principle, the CCM also in that case faces the problems mentioned above, like static kink-impurity forces or the fact that, in general, the kink-impurity mode does not factorize into a kink mode and an impurity mode. Nevertherless, it turns out that the CCM describes the numerical scattering results reasonably well. The spectral wall phenomenon was not discussed in that publication. 

{\bf Acknowledgements: \; }
The authors acknowledge financial support from the Ministry of Education, Culture, and Sports, Spain (Grant No. FPA2017-83814-P), the Xunta de Galicia (Grant No. INCITE09.296.035PR and Conselleria de Educacion), the Spanish Consolider-Ingenio 2010 Programme CPAN (CSD2007-00042), Maria de Maetzu Unit of Excellence MDM-2016-0692, and FEDER. Further, we thank H. Weigel for helpful remarks.
\bibliographystyle{apsrev4-1}
%

\end{document}